\documentclass[twocolumn]{aastex63}
\pdfoutput=1
\usepackage{xcolor}
\usepackage{makecell}
\usepackage{pbox}
\usepackage{gensymb}
\usepackage{float}

\received{February 21, 2021}
\accepted{July 26, 2021}

\shorttitle{Turbulence in the CND}
\shortauthors{Dinh et al.}

\usepackage{xcolor}

\newcommand{\x}{\color{blue}}

\newcommand{\msun}{M$_{\odot}$}
\newcommand{\gadget}{Gadget2}
\newcommand{\e}[1]{$\times10^{#1}$}

\newcommand{\invcmcube}{cm$^{-3}$}

\newcommand{\nt}{$N_t$}
\newcommand{\eturb}{$\Delta E_{in}$}
\newcommand{\splash}{\texttt{SPLASH}}

\begin{document}

\title{Effects of turbulence in the Circumnuclear Disk}

\correspondingauthor{Cuc~K.~Dinh}
\email{cdinh00@g.ucla.edu}

\author{Cuc~K.~Dinh}
\affiliation{Dept. of Physics \& Astronomy, University of California, Los Angeles, CA, 90095, USA}

\author{Jesus~M.~Salas}
\affiliation{Dept. of Physics \& Astronomy, University of California, Los Angeles, CA, 90095, USA}

\author{Mark~R.~Morris}
\affiliation{Dept. of Physics \& Astronomy, University of California, Los Angeles, CA, 90095, USA}

\author{Smadar~Naoz}
\affiliation{Dept. of Physics \& Astronomy, University of California, Los Angeles, CA, 90095, USA}
\affiliation{Mani L. Bhaumik Institute for Theoretical Physics, Department of Physics and Astronomy, University of California, Los Angeles, CA 90095, USA}
	
\begin{abstract}
A Circumnuclear Disk (CND) of molecular gas occupies the central few parsecs of the Galactic Center. It is likely subject to turbulent disruptions from violent events in its surrounding environment, but the effect of such perturbations has not yet been investigated in detail. Here we perform 3D, N-body/smoothed particle hydrodynamic (SPH) simulations with an adapted general turbulence driving method to investigate the CND's structural evolution, in particular its reaction to varied scales of injected turbulence. We find that, because of shear flow in the disk, transient arcs of gas (streams) naturally arise when turbulence is driven on large scales (up to $\sim4$~pc), as might occur when a supernova blast wave encounters the CND. Because energetic events arise naturally and often in the central parsecs of our Galaxy, this result suggests that the transient structures that characterize the CND do not imply that the CND itself is a transient structure. We also note that features similar to the density concentrations, or ``clumps'', detailed in literature emerge when we account for the observed orientation of the disk and for the spatial resolution of observations. As such, clumps could be an artifact of observational limitations.

\vspace{1cm}

\end{abstract}
\section{Introduction}
The Galactic Center (GC) is a dynamically complex region hosting a $4\times10^6$ \msun\ supermassive black hole manifested as the radio source Sgr A* \citep[e.g.,][]{Ghez2008}. The black hole is orbited by a ring of molecular gas known as the Circumnuclear Disk (CND, \citealt{Becklin1982ApJ...258..135B,Morris1996}). This moderately dense ($10^{4}-10^{6}$ cm$^{-3}$) and massive (few $10^{4}$ \msun) disk has an inner radius of $\sim1.5$~pc and a spatially variable outer radius of $3$-$7$ pc \citep{serabyn+86,Sutton+90,Morris1996,Oka+11,Etxaluze+11}. It is also relatively warm, with a temperature of a few hundred Kelvin \citep[e.g.][]{Genzel1985ApJ...297..766G,Guesten1987ApJ...318..124G,Genzel1989IAUS..136..393G, Jackson1993ApJ...402..173J,Morris1996,Bradford2005ApJ...623..866B,Requena-Torres2012A&A...542L..21R,Lau2013ApJ...775...37L,Mills2013ApJ...779...47M}, and a smaller amount of gas exceeding 500 K \citep{Mills+17}. 

Numerous observational studies have identified various structural components of the CND. For example, \cite{Martin2012A&A...539A..29M} detail three kinematically distinct molecular structures having separate inclinations and peak rotation velocities. Other features that have been described include lobes, extensions, arcs, streamers and arms  \citep[e.g.,][]{Guesten1987ApJ...318..124G, Sanders1998MNRAS.294...35S, Christopher2005ApJ...622..346C, Montero-Castano2009ApJ...695.1477M, Hsieh2017ApJ...847....3H}.
Density clumps are also frequently mentioned in the literature, though their definition  varies depending on the study \citep[e.g.,][]{Guesten1987ApJ...318..124G,Sutton+90,Jackson1993ApJ...402..173J,Marr+93,Shukla2004ApJ...616..231S, Christopher2005ApJ...622..346C, Montero-Castano2009ApJ...695.1477M,Martin2012A&A...539A..29M, Lau2013ApJ...775...37L,SmithWardle14}. 

At present, the origin of the CND remains an open question; some studies favor the notion that this disk is a transient structure, while others suggest that it is quasi-stable and long-lived. The transient argument is largely based on the disk's non-uniformity: the presence of clumps and streams (e.g., \citealt{Genzel1985ApJ...297..766G,Christopher2005ApJ...622..346C,Montero-Castano2009ApJ...695.1477M,Martin2012A&A...539A..29M,Lau2013ApJ...775...37L,Hsieh+21}).  Because the rotational dynamics of the disk are dominated almost entirely by the central black hole, the disk should experience strong differential rotation that smooths out non-uniformities such as clumps on an orbital time scale. The observed non-axisymmetric density distribution has therefore been invoked as evidence that either the CND was recently formed, and that the observed structure is therefore short-lived \citep[e.g.,][]{Guesten1987ApJ...318..124G, Requena-Torres2012A&A...542L..21R}, or that the structures within the disk are dense enough to be tidally stable \citep{Jackson1993ApJ...402..173J, Vollmer:2001a, Vollmer:2001b, Shukla2004ApJ...616..231S, Christopher2005ApJ...622..346C, Montero-Castano2009ApJ...695.1477M,Hsieh+21}. However, multiple studies have  found that the clumps have densities ranging up to a few times $\sim10^{5}$ cm$^{-3}$, which falls well below the critical Roche density of $\sim3\times10^{7}$ cm$^{-3}$ (e.g., \citealt{Guesten1987ApJ...318..124G,Requena-Torres2012A&A...542L..21R, Lau2013ApJ...775...37L,Mills2013ApJ...779...47M}). Thus, the observed density concentrations are tidally unstable, which has led some to conclude that the CND is a transient structure, with a lifetime of a few dynamical timescales ($\sim10^{4}$-$10^{5}$~yrs, \citealt{Guesten1987ApJ...318..124G}). A popular model to account for the transient features of the CND is that a passing cloud was gravitationally captured by the black hole. This ``infalling cloud'' quickly circularized as it was tidally disrupted into what is now the CND \citep{Sanders1998MNRAS.294...35S, Bradford2005ApJ...623..866B, Bonnel2008Sci...321.1060B, Wardle2008ApJ...683L..37W, HobbsNayakshin09,  Alig2011MNRAS.412..469A, Oka+11,  Mapelli2012ApJ...749..168M, Mapelli+Trani16, Trani2018ApJ...864...17T, Goicoechea+18a,  Ballone+19}.  

 The long-term evolution of the CND is therefore the question before us.  Are we seeing it soon after it has been created, and will it persist as a quasi-stable structure well into the future, or will it be dissipated by various forms of energetic activity at the Galactic center? An important factor to consider when addressing these questions is that the CND is subject to various sources of turbulence, which come from the active environment of the Galactic Center. These can be in the form of supernovae (e.g., \citealt{Mezger1989A&A...209..337M, Martin2012A&A...539A..29M}), outflows from the Nuclear Stellar Cluster (e.g., \citealt{Morris1996,Genzel2010RvMP...82.3121G}) and Sgr A*, and turbulence due to the CND's toroidal magnetic field, which has a strength on the order of a few milligauss (e.g., \citealt{Marshall+95, Bradford2005ApJ...623..866B,Hsieh2018ApJ...862..150H}). These perturbers may play a vital role in the CND's morphology. However, the general effect of turbulence driving on the disk's structural evolution is still largely uncertain.

In this paper, we perform 3D smoothed particle hydrodynamic (SPH) simulations that include self-gravity and a turbulence driving method to investigate how different scales of turbulence affect the CND's dynamical evolution. Our aim is to determine whether turbulence can help us distinguish between the transient and non-transient hypotheses. This paper is organized as follows: In Section \ref{sec:numerical_methods} we describe the numerical code and methods employed. In Section \ref{sec:results} we present our simulation results.  Finally, we summarize our paper and offer some insights spawned by our results in Section \ref{sec:discussion}, and compare our results with observations, noting the similar morphological features.

\section{Numerical Methods}\label{sec:numerical_methods}
We used the N-body/SPH code \gadget\ \citep{Springel2005}, which is based on the tree-Particle Mesh method for computing gravitational forces and on the SPH method for solving the Euler equations of hydrodynamics. The smoothing length of each particle in the gas is fully adaptive down to a set minimum of 0.001 pc. \gadget\ employs an entropy formulation of SPH, as outlined in \cite{Springel2002}, with the smoothing lengths defined to ensure a fixed mass (i.e., fixed number of particles) within the smoothing kernel volume (set at N$_{neigh}$ = 64). The code adopts the Monaghan-Balsara form of artificial viscosity \citep{Monaghan1983,Balsara1995}, which is regulated by the parameter $\alpha_{MB}$, set to $0.75$.

We modified the standard version of \gadget\ to include turbulence driving, the gravitational potential of the inner $10$ pc of the Galaxy, and the effects of  pressure by a surrounding medium. We describe these modifications below.

\subsection{Gravitational potential}\label{subsec:potential}
We include into Gadget2 the gravitational potential
of the inner $10$ pc of the Galaxy, which includes the supermassive black hole (Sgr A*) and the Nuclear Stellar Cluster (NSC, e.g., \citealt{Do2009ApJ...703.1323D,Schodel2009A&A...502...91S}). In other words, the total potential, $\Phi_{T}$ is: 
\begin{equation}
  \Phi_{T}=\Phi_{bh}+  \Phi_{NSC} \ .
\end{equation}
The gravitational potential of the black hole is included as a point-mass potential:
\begin{equation}
\Phi_{bh} = -\frac{GM_{bh}}{r} \ ,
\end{equation}
where $M_{bh}=4\times10^6$ \msun, and $r$ is the radial distance from SgrA*.
Furthermore, to include the gravitational effects of the NSC, we adopt the potential described by \cite{Stolte2008}:
\begin{equation}\label{eq:nsc}
    \Phi_{NSC} = \frac{1}{2} v_c^2 \ln{(R_c^2 + r^2)} \ , 
\end{equation}
where $v_c = 98.6$~km/s is the velocity on the flattened portion of the rotation curve at very large radii, and $R_c = 2$~pc is the core radius.

\subsection{External pressure}\label{subsec:pressure}
The interstellar medium surrounding the CND is modeled via an external pressure term to approximate a constant pressure boundary. Following \cite{Clark2011}, we modify \gadget's momentum equation \citep{Springel2002}:
\begin{equation}\label{eq:sph_motion2}
\frac{d v_i}{dt} = - \sum_{j} m_j \left[  f_i \frac{P_i}{\rho^2_i} \nabla_i W_{ij}(h_i) + f_j \frac{P_j}{\rho^2_j} \nabla_i W_{ij}(h_j)  \right] \ ,
\end{equation} 
where $v_i$ is the velocity of particle $i$, $m_j$ is the mass of particle $j$, $P_i$ is the pressure, $\rho_i$ is the density, $W_{ij}(h_i)$ is the kernel function which depends on the smoothing length $h_i$,  and the coefficients $f_i$ are defined as $f_i = \left( 1 + \frac{h_i}{3\rho_i} \frac{\partial \rho_i}{\partial h_i} \right)^{-1}$. We replace $P_i$ and $P_j$ with $P_i - P_{ext}$ and $P_j - P_{ext}$, respectively, where $P_{ext}$ is the external pressure. The pair-wise nature of the force summation over the SPH neighbors ensures that $P_{ext}$ cancels for particles that are surrounded by other particles. At the boundary, where the $P_{ext}$ term does not disappear, it mimics the pressure contribution from a surrounding medium \citep{Clark2011}. We set $P_{ext}$ equal to $10^{-10}$ ergs \invcmcube, an approximate value for the GC \citep{Spergel1992,Morris1996}.

\subsection{Initial conditions}\label{subsec:IC}

For simplicity, we start with a gas disk having initial parameters drawn from observations: an inner radius of $1.5$ pc, outer radius $4.5$ pc, and a mass of $4.5\times10^4$ \msun. The disk contains $10^6$ SPH particles, and the initial density is uniform in the disk midplane with a value of $\approx 10^{5}$ cm$^{-3}$, and has a Gaussian vertical distribution with a scale height of 0.2 pc.
The particles are initially in circular orbits, with their velocities calculated using the potential described in Section \ref{subsec:potential}. All simulations were run using an isothermal equation of state with $T$ = $200$ K. 

\subsection{Turbulence driving}\label{subsec:turbulence}
We use the turbulence driving method described by \cite{Salas2021AJ....161..243S}, which consists of a Fourier forcing module modelled with a spatially static pattern. This turbulence module follows the methods by \cite{Stone1998ApJ...508L..99S} and \cite{MacLow1999}. For completeness, we recapitulate the key factors of the algorithm here.

First, we create a library of $10$ files of turbulence, which our modified version of the Gadget2 code reads in at the start of  the simulation. Each file contains a unique realization of a turbulent velocity field (in the form of a 3D matrix) with power spectrum $P(k) \propto k^{-4}$ (where $k$ is the wavenumber). This power law is steeper than Kolmogorov turbulence, but it is suitable for compressible gas \citep{Clark2011}. Each 3D matrix is created using fast Fourier transforms inside a $128^3$ box, following the techniques described in \cite{Rogallo1981} and \cite{Dubinski1995}. These 3D turbulence matrices can be visualized as lattice cubes (or grids) containing $128\times128\times128$ equally spaced lattice points. We set the physical size of these cubes to $L_{cube}$ per side, depending on the turbulence model (see Section \ref{subsec:models}).  

In some studies, the driving module only contains power on the larger scales (e.g., \citealt{Federrath2010AA...512A..81F}). This type of driving models the kinetic energy input from large-scale turbulent fluctuations, which then break up into smaller structures as the kinetic energy cascades down to scales smaller than the turbulence injection scale. However, in SPH, the artificial viscosity can damp this energy cascade and prevent it from reaching the smaller scales. Consequently, to create the different realizations of turbulent velocity fields, we use a discrete range of $k$ values from $k_{min} = 2$ to $k_{max} = 128$, thus effectively injecting energy on scales between $\lambda_{max} = L_{cube}/2$ and $\lambda_{min} = L_{cube}/128$. 

Then, the volume of our simulation domain ($\sim10$ pc per side) is filled with a number $N$ of these cubic lattices per side (where $N = 4$ or $100$, depending on the model, see Section \ref{subsec:models}), each drawn randomly from our library. Using this method, the spatial resolution of the turbulence in our simulation domain is the same as the resolution of an individual turbulence lattice cube. 

To drive the turbulence, we follow a method similar to that described by \cite{MacLow1999}: every \nt\ timesteps (the timestep in all our simulations is fixed to $\Delta t_s = 100$ yrs) we add a velocity increment to every SPH particle, $i$, given by:
\begin{equation}
\label{eq:kicks}
\Delta \vec{v}_i(x,y,z) = F(\rho_i) A \vec{I}_i(x,y,z)\ ,
\end{equation}
where $\vec{I}_i(x,y,z)$ is the turbulent velocity interpolated from the turbulence field of the lattice cube that contains the particle in question. The amplitude $A$ is chosen to maintain a constant kinetic energy input rate $\dot{E_{in}} = \Delta E_{in}/(N_t \Delta t_s)$, and the term $F(\rho_i)$ is a factor that depends on the particle's density, $\rho_i$, and on the turbulence model (see Section \ref{subsec:models}). Any particle outside the simulation domain does not receive any turbulent energy. 

Our turbulence implementation contains two free parameters: \eturb, the total energy input per instance of turbulent energy injection, and  $\Delta t = N_t \Delta t_s$, the time between velocity ``kicks''. We show below the chosen parameters in our models.

\subsection{Turbulence models}\label{subsec:models}

\begin{table}
\begin{center}
\begin{tabular}{|c|c|c|c|}
\hline
Run name  & Turbulence scale & \eturb\ (ergs) & $\Delta t$ (yrs) \\ \hline
STS-t100-47 & Small  & $10^{47}$  & 100  \\ 
        & (0.05-10$^{-4}$~pc)   &     &   \\
\hline
STS-t500-47 & Small     & $10^{47}$ & 500   \\
        & (0.05-10$^{-4}$~pc)   &     &   \\
\hline
LTS-t$5$-$49$    & Large     & $10^{49}$    & $10^5$  \\ 
        & (2-0.3 pc)   &     &   \\
\hline
LTS-t$5$-$50$    & Large      & $10^{50}$ & $10^5$   \\ 
        & (2-0.3 pc)   &     &   \\
\hline
\end{tabular}
\caption{Parameters of all turbulence runs. \eturb\ corresponds to the total energy per injection, and $\Delta t$ = \nt $\Delta t_s$ is the time between injections. }
\label{table:models}
\end{center}
\end{table}

Here we describe our turbulence models and assumptions. We consider $1$) a large-scale model and $2$) a small-scale model, each corresponding to different physical scenarios. Table \ref{table:models} summarizes the runs and parameters to be used based on our models.

\subsubsection*{Model $1$ - Small turbulence scales (STS)}
We first investigate a small-turbulence-scale model by setting $L_{cube}$ = $0.1$ pc. Thus, energy is injected on scales $\lambda = 0.05$~pc to $10^{-4}$~pc (for $k=2$ and $k=128$, respectively). We fill the volume of our simulation domain ($10\times10\times10$~pc) with $N=100$ of these boxes per side. This model can represent, for example, turbulence driven by stellar winds, occasional novae, MHD instabilities, outflows from Wolf-Rayet stars throughout the CND, and any other source of turbulence operating on small scales. We expect such perturbations to affect the surrounding gas of the CND on few-hundred-year timescales.

Therefore, we examine two injection timescales, $\Delta t$ = $100$~yrs and $\Delta t$ = $500$~yrs, and an injection energy of $10^{47}$~ergs. The physical processes mentioned above are not expected to exceed these energy rates. Moreover, as we report in Section \ref{sec:results}, the injected turbulence has little to no significant effect on the model's morphological evolution, except to progressively thicken the disk as the energy increases.

Furthermore, in \cite{Salas2021AJ....161..243S} the turbulent velocity kicks have a density dependence factor $F(\rho) = \sqrt{G \rho}$, which helps counteract the effects of run away self-gravity by preventing the unphysical formation of extremely high density clumps, which can happen in an SPH formulation on scales similar to the smoothing length  (c.f, Section $3.2$). This density factor is not needed in this STS model since a temperature of $200$~K creates a sufficient thermal pressure gradient in the gas disk to counteract its self-gravity. Thus, in this model we eliminate this density dependence by setting $F(\rho) = 1$.

\begin{figure*}
\centering
\includegraphics[width=1\textwidth]{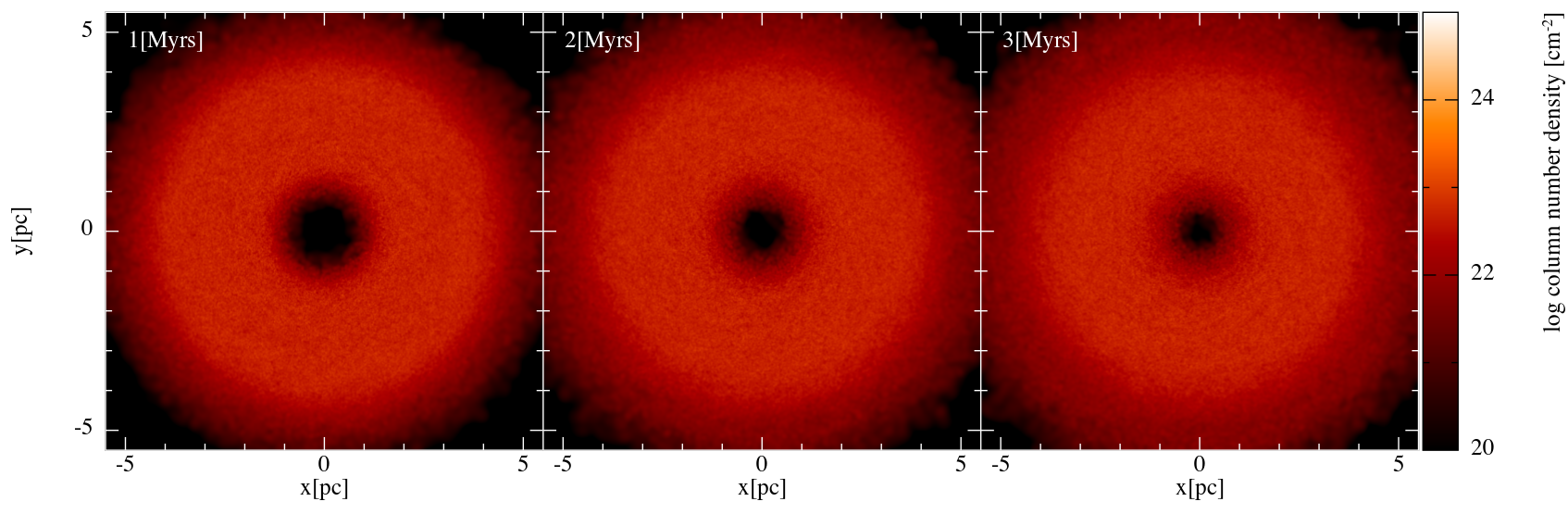}
\includegraphics[width=1\textwidth]{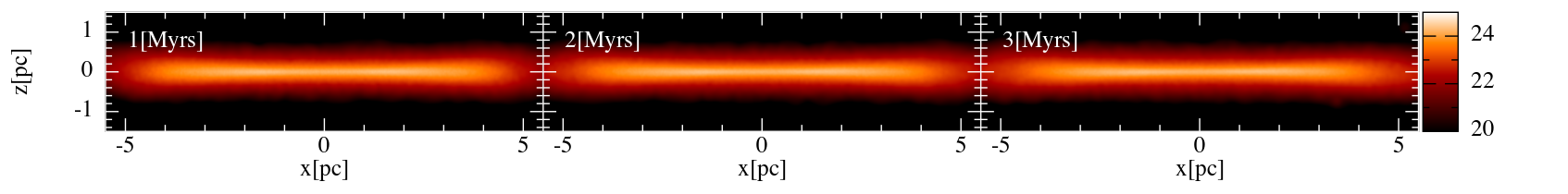}
\caption{Column number density of the Small Turbulence Scale (STS) model featuring its evolution through the simulation. The simulation was run for 3 Myrs with injection energy of 10$^{47}$ ergs at intervals of $\Delta t_N$ = 100 years. Top panel: Face-on view; bottom panel: edge-on view.}
\label{fig:STS_aden}
\end{figure*}
\subsection*{Model 2 - Large turbulence scales (LTS) }
We next consider a large-turbulence-scale model by setting $L_{cube}$ = $4$~pc. Thus, energy is injected on scales $\lambda = 2 - 0.03$ pc (for $k=2$ and $k=128$, respectively). This model represents, for example, turbulence driven by supernovae (SNe). We fill the volume of our simulation domain with $N=4$ of these boxes, each box occupying a quadrant of the simulation domain (which in this case has a volume of $8\times8\times4$ pc). Because the effect of a supernova is localized to a few parsecs and not distributed uniformly across the disk, we activate the turbulence driving only in one randomly chosen quadrant of the simulation volume at a time. 

We expect disturbances caused by SNe to drive turbulence over longer timescales than in the case of STS. Thus, we adopted an injection interval of $10^{5}$ yrs, roughly the average lifetime of a supernova remnant. Furthermore, given the number of massive stars within the central few parsecs of the Galaxy (e.g., \citealt{Do+13a,Nishiyama2013A&A...549A..57N}) we might expect their recurrence timescale to be also on the order of $10^{5}$ years. SN explosions release $\sim10^{51}$~ergs of energy, which drives a blast wave through the ambient ISM. However, the properties of the environment within which a SN explodes strongly affect the fraction of energy that is deposited into the surrounding gas \citep{Dwarkadas2012MNRAS.419.1515D}. For example, \cite{Cox1972ApJ...178..159C} estimates that $10$\% of the initial SN energy is retained as thermal and kinetic energy in the gas.  \cite{Chevalier1974ApJ...188..501C} obtained a fraction between $4$-$8$\%, depending on the ambient density, assumed homogeneous. Other studies have found fractions from up to $50$\% \citep{McKee1977ApJ...218..148M,Cowie1981ApJ...247..908C} to just a few percent \citep{Slavin1992ApJ...392..131S,Walch2015MNRAS.451.2757W}. Therefore, we opted for testing energy injection values of \eturb = $10^{49}$ and $10^{50}$ ergs, which correspond to 1\% and 10\% of a typical SN energy, respectively. Additionally, here we use a density factor of $F(\rho) = \rho^{-1/4}$ (see Equation \ref{eq:kicks}) to capture the density dependence of the expansion velocity of a SN shockwave during the snowplow phase \citep[e.g.,][]{Cioffi1988ApJ...334..252C}. 

Finally, as was shown in \cite{Salas2021AJ....161..243S}, our turbulence method enhances the inward angular momentum transport of gas. In this model, rather than a slow inward migration, the large-scale perturbations could potentially send gas onto ballistic trajectories towards the center. Thus, to counteract the rapid filling of the CND cavity due to turbulence, we mimic the outflow from the young nuclear cluster of massive, windy stars. We achieve this by adding a radially outwards velocity, $v_i(r)$, to every SPH particle, $i$, within $1.5$~pc (see Appendix \ref{appen:wind} for further details):
\begin{equation}
v_i(r) =  \zeta v_{esc,i}(r) \sqrt {\frac{n_0}{n_i}} \ ,
\end{equation}
where $v_{esc,i}(r)$ is the escape velocity of an SPH particle at radius $r$, $n_0 = 10$ \invcmcube\ is the assumed number density of the wind where it strikes the inner edge of the CND (the simulations by \citealt{Blank2016MNRAS.459.1721B} assumed an outflow density of $n_0=100$ \invcmcube\ at $r=0.5$~pc. However, since the outflow's density decreases by $1/r^2$, $n_0\approx10$  \invcmcube\ at $r=1.5$~pc), and $n_i$ is the number density of an SPH particle at the inner edge of the CND. Finally, we add the free parameter $\zeta$ in order to adjust the magnitude of this radial velocity so that the inner edge maintains a stable radius. This also avoids launching particles off to infinity, which creates numerical errors in the simulations. In our calculations, $\zeta = 0.15$. 

We note that this treatment is artificial, since a realistic treatment of the NSC's outflow proves to be extremely difficult with Gadget2. However, our goal here is only to prevent the rapid inward migration provoked by the turbulence, which causes numerical problems when particles pile up near the origin.

\section{Results}\label{sec:results}

\begin{figure*}
\centering
\includegraphics[width=1\textwidth]{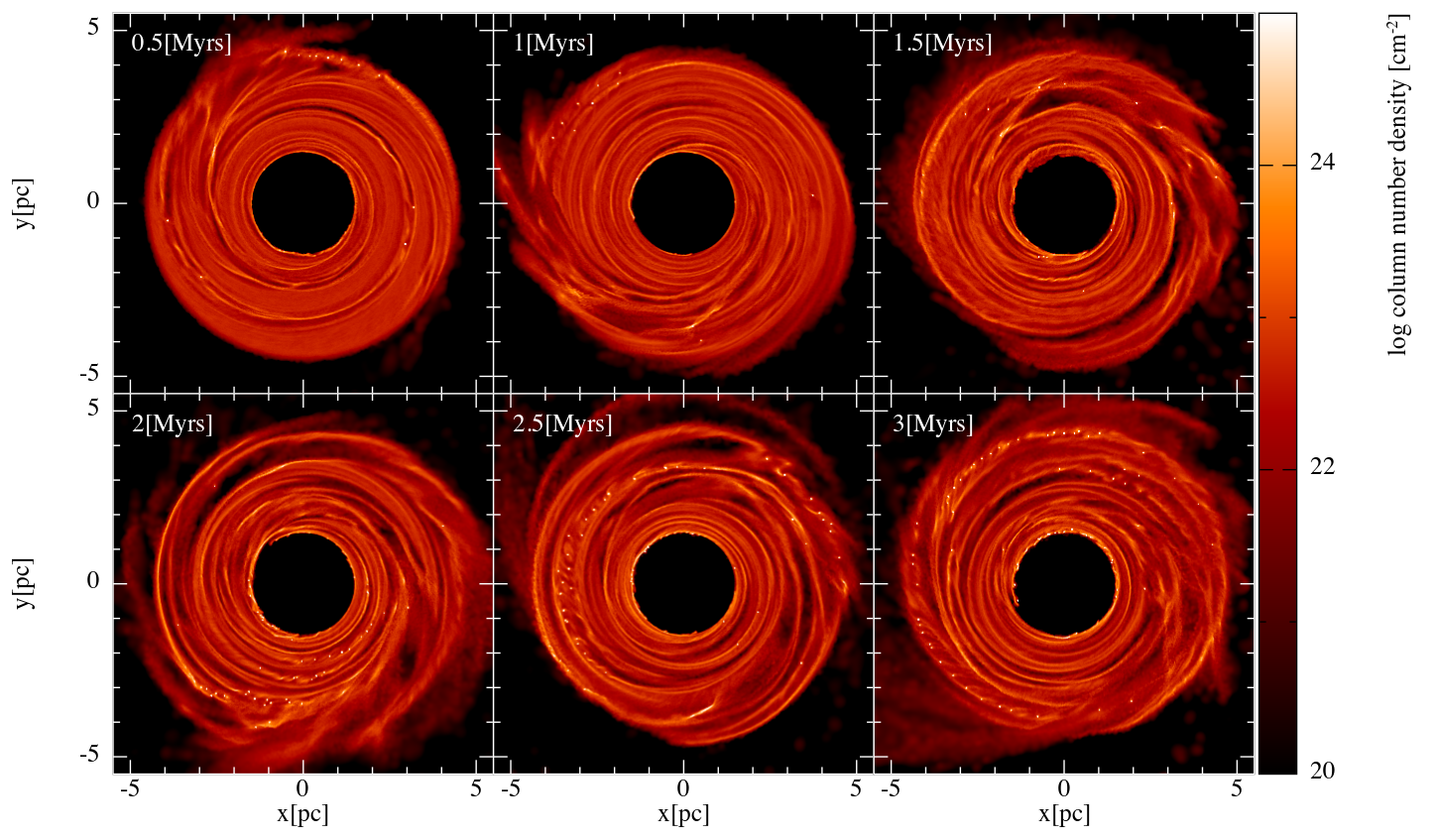}
\includegraphics[width=0.98\textwidth]{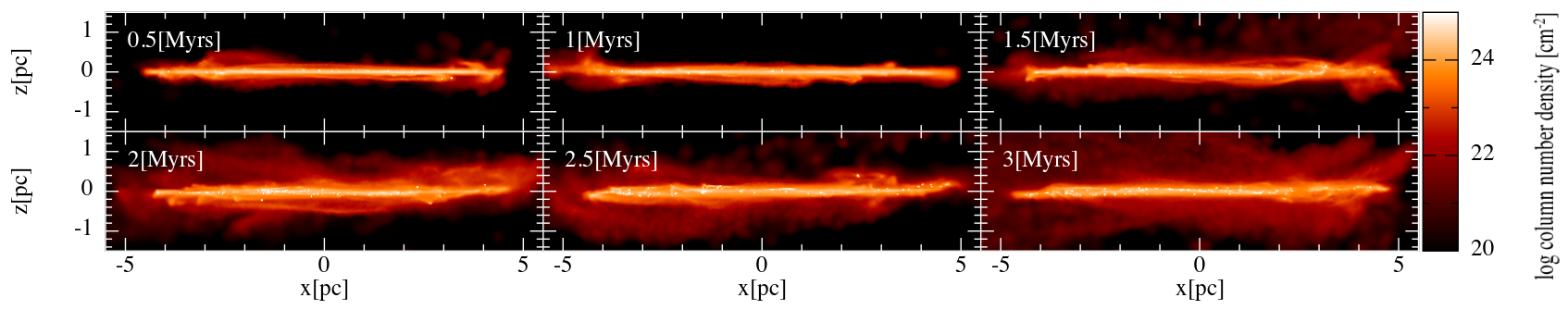}
\caption{Column number density maps featuring the evolution of the Large Scale Turbulence model (LTS-t$5$-$49$). The run was run for 3 Myrs with injection energy of $10^{49}$ ergs at intervals of $\Delta t$ = $10^{5}$ years. Top: face-on view, rotates counter-clockwise. The disk is fully and densely permeated by identifiable spiral streams by $\sim1$~Myr after the start of the simulation. Compact, high-density concentrations appear by $0.5$ Myrs but are not prominent until $2$ Myrs. Bottom: edge-on view. Material is increasingly perturbed off the disk as the model evolves.}
\label{fig:LTS_t549_aden_evol}
\end{figure*}

\begin{figure*}
\centering
\includegraphics[width=1\textwidth]{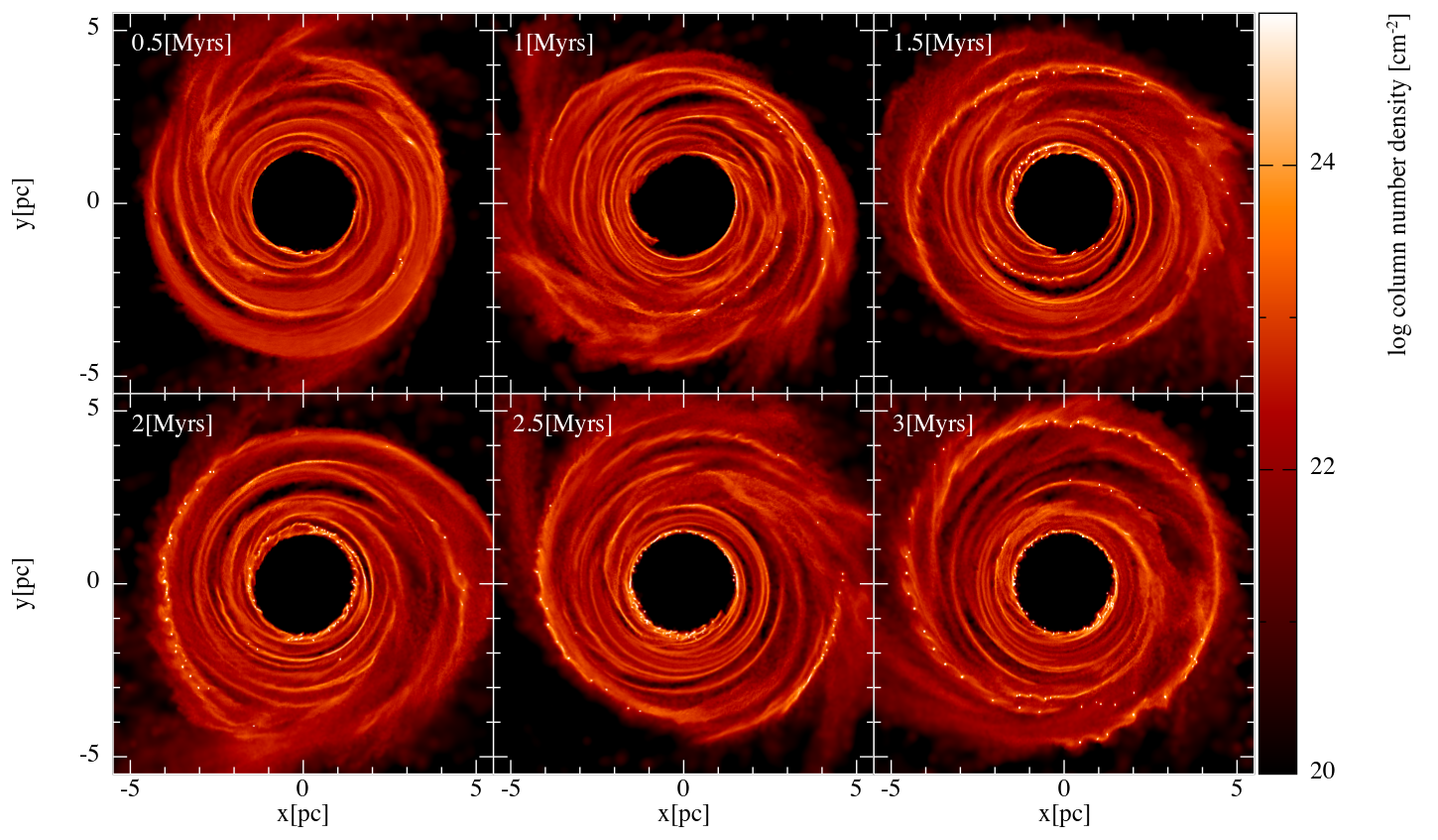}
\includegraphics[width=1\textwidth]{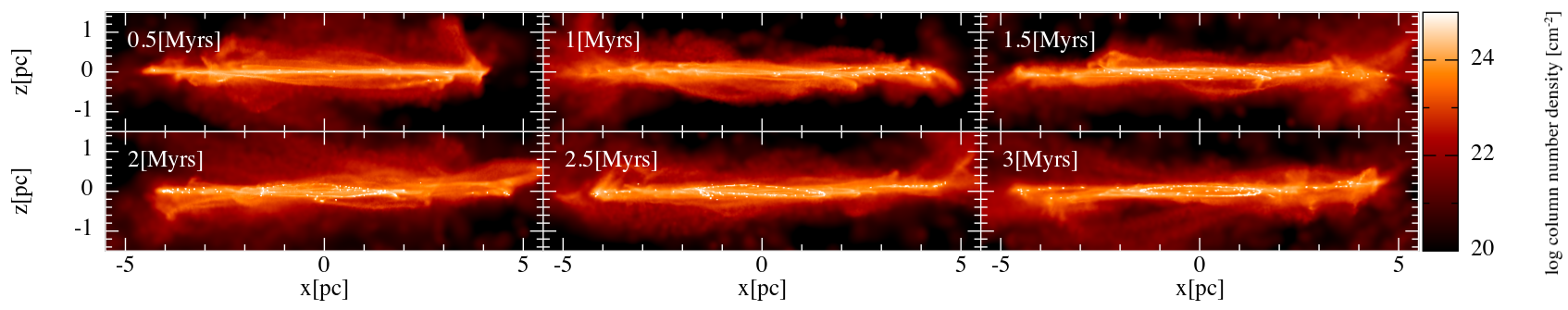}
\caption{Column number density maps depicting the evolution of the Large Turbulence Scale model (LTS-t$5$-$50$). The simulation was run for $3$ Myrs with injection energy of $10^{50}$ ergs at intervals of $\Delta t = 10^{5}$ years. Top: face-on view, rotates counter-clockwise. Much more distinct streams with higher densities (especially comparing to LTS-t$5$-$49$ at $3$ Myrs). Bottom: edge-on view. More prominent perturbations off the disk plane than in the LTS-t$5$-$49$ model, as expected. }
\label{fig:LTS_t550_aden_evol}
\end{figure*}

\begin{figure*}
\centering
\includegraphics[width=0.98\textwidth]{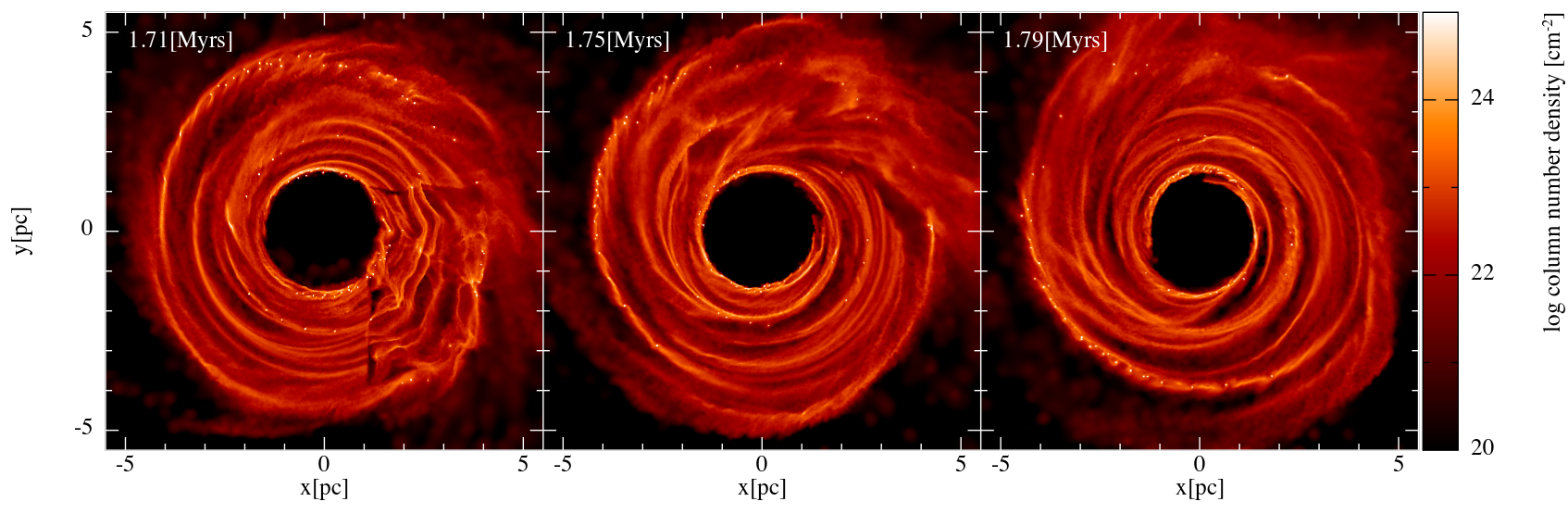}
\caption{Column Number density map depicting the evolution of the Large Turbulent Scale model (LTS-t$5$-$50$) between two energy injections. Face-on view with the disk rotating counterclockwise. The imprint of the turbulence grid just after it is imposed is apparent at $1.71$ Myrs (left panel). While the geometry of the turbulence injection is physically oversimplified, this structure is quickly dispersed by differential rotation, forming orbiting spiral streams that would undoubtedly be also produced by a more spatially continuous injection prescription. } 
\label{fig:LTS_t550_inject}
\end{figure*}

\begin{figure*}
    \centering
    \includegraphics[width=0.95\textwidth]{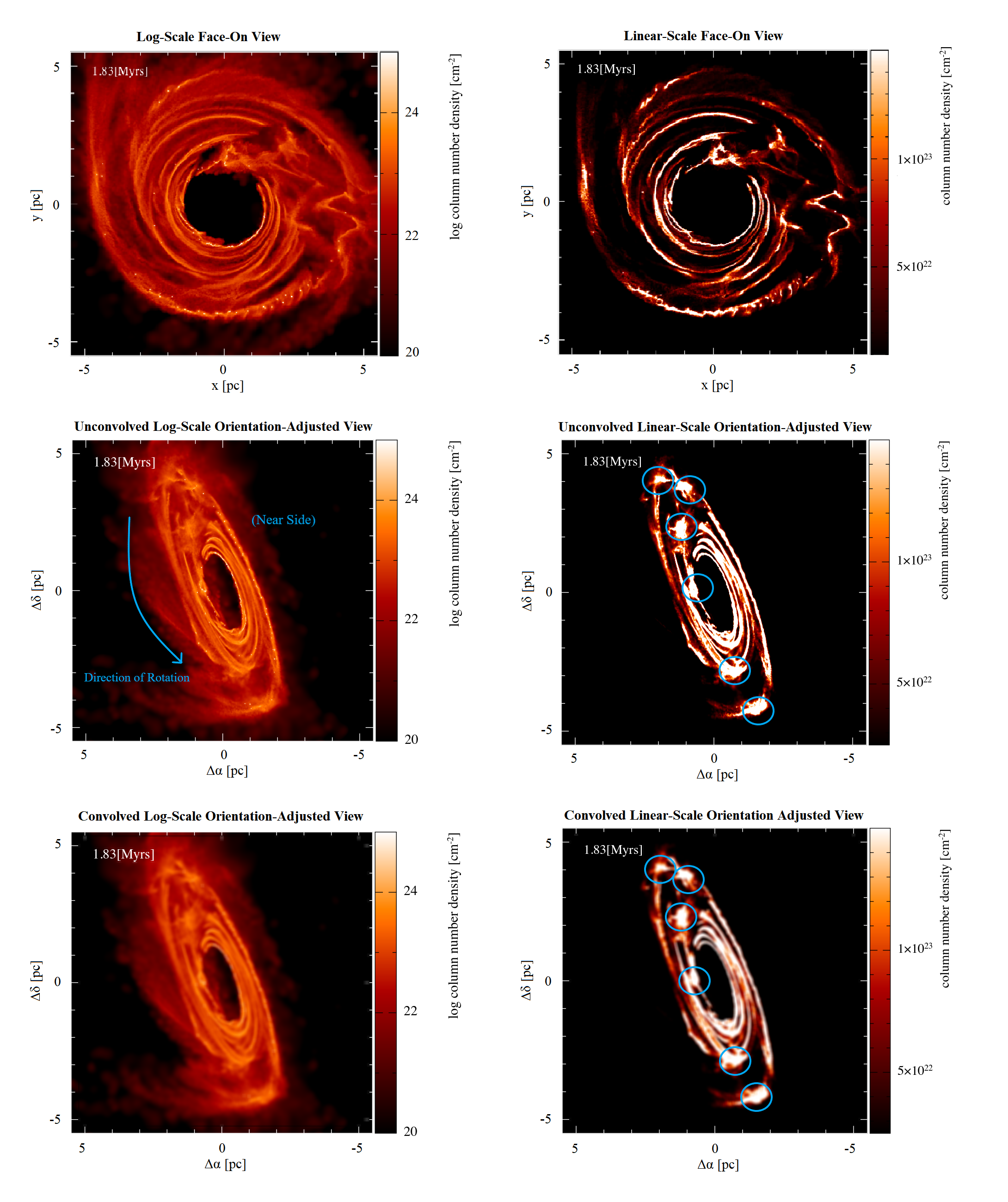}
    \caption{Column number density maps of the LTS-t$5$-$50$ model (domain size of $11\times11$ pc$^2$) at $1.83$~Myrs, compared between linear scale (right panels) and log scale (left panels). The upper two panels show the face-on log-scale column number density map and its linear scale counterpart. The middle panels account for the projection of the disk, and the bottom panels are the corresponding convolved counterparts. The identified clumps have been circled in blue in both the unconvolved (middle right) and convolved (bottom right) linear-scale orientation-adjusted view. These clumps seem in part to be a confluence of distorted streams.}
    \label{fig:convolve}
\end{figure*}

We  tested each turbulence run for $3$ Myrs, which is far greater than the dynamical timescale of the disk ($\sim0.1-0.3$~Myrs), allowing us to assess the long-term effect of the induced perturbations. The results are shown and described below.

\subsection{Model 1 (STS)}

Snapshots of the time evolution of the STS-t$100$-$47$ run are shown in Figure \ref{fig:STS_aden}, where we show column number density maps at $1$, $2$ and $3$ Myrs (we present only one of the runs, since all STS tests are qualitatively identical). Given that the spatial scales over which the energy injection is distributed are relatively small, the net effect of turbulence is manifested on only the smallest of scales, as expected. The small-scale converging flows create a granular structure throughout the disk, and this texture continues during the entire simulation. 

Furthermore, our driving method creates a ``turbulent'' viscosity which promotes the transfer of angular momentum, as discussed in \cite{Salas2021AJ....161..243S}. These viscous forces broaden the disk slightly: the disk expands radially outwards, which increases the outer radius, while the inner radius decreases slightly. The cavity closes at a faster rate than the disk's expansion, but the inward migration rate was still relatively slow. We did not impose an artificial radial outflow as in the LTS runs, since we presumed, correctly, that it was not needed. Overall, the STS models reached steady state early on in the simulations. 

\subsection{Model 2 (LTS)}

The evolutionary progressions of the LTS-t$5$-$49$ and LTS-t$5$-$50$ runs are shown in Figures \ref{fig:LTS_t549_aden_evol} and \ref{fig:LTS_t550_aden_evol}, respectively. The effects of the turbulence are much more prominent here than in the STS models. The large-scale turbulence perturbations are quickly stretched out into long arcs by the strong tidal shear, which promotes the formation of continuous, orbiting spiral stream segments. We note that while these features are apparent in both runs, streams in the LTS-t$5$-$49$ run are less distinguishable in comparison to those in the LTS-t$5$-$50$ run. For example, the streams fully encapsulate the disk by $\sim{0.5-1}$~Myrs in the LTS-t$5$-$50$ run, but only by $\sim1.5$~Myrs in the LTS-t$5$-$49$ run. This is because the ten times greater energy in the LTS-t$5$-$50$ simulation provokes more prominent perturbations which are then sheared by the disk's rotation into higher contrast streams. This effect is showcased in Figure \ref{fig:LTS_t550_inject}, where we plot the column number density map of the LTS-t$5$-$50$ run between two energy injections. The imprint of the turbulence grid is apparent at $1.71$ Myrs (left panel in Figure \ref{fig:LTS_t550_inject}). This imprint occurs at every injection, and it was was an expected result, since the size of the turbulence grid is comparable to the scale of the CND. While the injection procedure in this LTS model is geometrically oversimplified, the grid's imprint, as well as the turbulent perturbations, disappear and are quickly sheared into orbiting streams, leaving no memory of the shape of the initial disturbance domain. 

The bottom panels of Figures \ref{fig:LTS_t549_aden_evol} and \ref{fig:LTS_t550_aden_evol} show that material was perturbed off the disk as the simulations evolved. This is because the turbulent velocity kicks in the LTS simulations are density-dependent: due to the imposed density factor of $F(\rho) = \rho^{-1/4}$ (see Section \ref{subsec:turbulence}), lower density promotes higher turbulent kicks. Thus, the turbulence is very effective in perturbing the lower density gas off the disk plane, creating an ``atmosphere'' of low-density gas surrounding the higher density disk. As such, the LTS-t$5$-$50$ run, due to its higher energy injection rate, had more material perturbed off the disk plane (see bottom panel of Figure \ref{fig:LTS_t550_aden_evol}) compared to the LTS-t$5$-$49$ run (bottom panel of Figure \ref{fig:LTS_t549_aden_evol}). In the sections that follow, we focus our analysis on the LTS-t$5$-$50$ model due to its more prominent structure.

We note the presence of compact, high-density concentrations (the small white points in Figures \ref{fig:LTS_t549_aden_evol} and \ref{fig:LTS_t550_aden_evol}) in both runs. These points formed early on and slowly increased in number over the course of the simulations. These clumps originate because here, local regions collapse under self-gravity. Particles pile on top of each other,  and due to  the nature of the kernel used by Gadget2, once the distance between particles approaches the smoothing length, the pressure gradient is no longer correct and the particles stick together, creating high-density concentrations. These clumps are, in effect, sink particles that might have registered as stars if the simulations had included a prescription for star formation. Many of these clumps originate in the inner rim of the disk, which is a consequence of the piling up of particles due to the artificial radial outflow. Thus, we ignore these clumps and consider them as numerical artifacts.  Their total mass is negligible, so they don't affect the evolution of the disk.  

\subsection{Comparing with observations: clumps}

In order to better compare our simulation with observations of the CND, we must account for 1) the orientation of the CND in the sky, and 2) the display function of the observational data, be it linear or logarithmic.

Thus, we orient the model disk to account for the observed orientation, with a disk inclination of $70^{\circ}$ and with the major axis at a position angle of $20^{\circ}$ 
\citep[e.g.,][]{Martin2012A&A...539A..29M, Lau2013ApJ...775...37L}. We also display this orientation-adjusted view with a linearly scaled color map. An example is shown in Figure \ref{fig:convolve} (middle right panel), which shows a snapshot of the LTS-t$5$-$50$ run at $1.83$~Myrs. In addition, we convolve this column density map of the tilted LTS-t$5$-$50$ model with a 2D Gaussian kernel to match the typical full width at half-maximum beam size of existing observations ($3.5$'', or $0.15$ pc, e.g., \citealt{Martin2012A&A...539A..29M, Lau2013ApJ...775...37L, Hsieh2017ApJ...847....3H}). Adjusting the orientation of the model (shown in the middle right panel of Figure \ref{fig:convolve}) gives rise to density enhancements. These enhancements, or knots, appear as clumps when the distorted streams are fortuitously projected. They are further realized as the convolutions smooth out the model. However, it is important to note that these clumps are still part of the larger stream-like structures, and that most of them are not identifiable as such in the face-on view of the simulation (top row of Figure \ref{fig:convolve}).

The sizes ($\sim0.3-0.5$~pc) and densities ($1.3 \times 10^{5}$ to $1.5 \times 10^{5}$ cm$^{-3}$) of these apparent clumps seem to be comparable to those described in the literature \citep{Guesten1987ApJ...318..124G,Sutton+90,Jackson1993ApJ...402..173J,Marr+93,Shukla2004ApJ...616..231S, Christopher2005ApJ...622..346C, Montero-Castano2009ApJ...695.1477M,Martin2012A&A...539A..29M, Lau2013ApJ...775...37L,SmithWardle14}, though the term ``clumps'' is defined in a variety of ways. For example, \citet{Lau2013ApJ...775...37L} describe clumps characterized by enhanced density at the inner edge of the CND. These are possibly caused by Rayleigh-Taylor instabilities derived from the interaction of the central stellar winds and the inner edge of the CND \citep[e.g.,][]{Blank2016MNRAS.459.1721B}. On the other hand, \cite{Montero-Castano2009ApJ...695.1477M} describe the CND to be composed of tidally stable blobs with varied sizes ($\sim{0.15}$-${0.54}$ pc), densities ($\sim{10^{7}}$ to $10^{8}$ cm$^{-3}$), and radial distances from Sgr A*. 

In our model, however, apparent clumps emerge as a result of the disk projection, coupled with sampling with finite spatial resolution. In addition, when considering the two presentations of the orientation-adjusted model (middle and lower right panels of Figure \ref{fig:convolve}), we find the clumps to coincide with the densest sections of the streams. This result leads us to pose the question of whether at least some of the clumps seen in observational studies might be projections of turbulence-induced, transient unresolved structures in the CND. Perhaps further investigation of the CND’s internal structure with higher resolution instruments such as ALMA could settle this question. 

\subsection{Comparing with observations: streams}

\begin{figure*}[!t]
    \centering
    \includegraphics[width=0.8\textwidth]{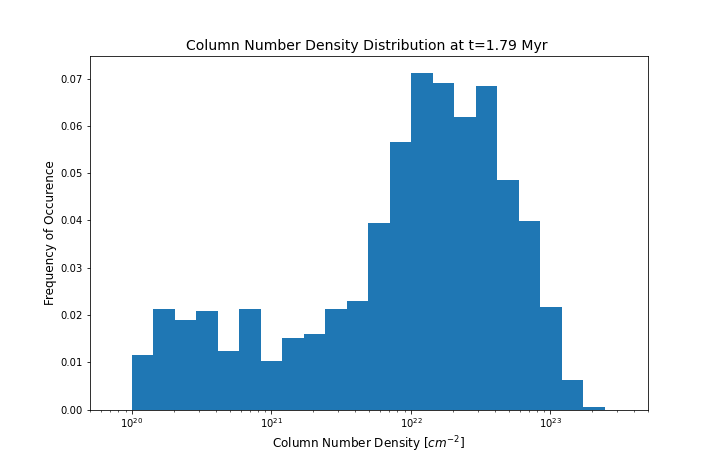}
    \caption{The column number density distribution of the LTS-t$5$-$50$ model (\eturb = $10^{50}$ ergs with $\Delta t$ = 10$^{5}$ years) at $1.79$ Myrs.}
    \label{fig: density_distribution}
\end{figure*}

The intrinsic structure in our LTS models consists of tidally stretched streams, which are evident when inspecting the column number density maps (see Figures \ref{fig:LTS_t550_aden_evol} and \ref{fig:convolve}), especially the linearly scaled maps (e.g., right-hand panels of Figure \ref{fig:convolve}). These streams have a high density contrast compared to the inter-stream regions: by $1.79$ Myrs, the column number density varies by roughly two orders of magnitude between $1.65$ and $7.65$ pc from the center (see Figure~\ref{fig: density_distribution}) with a median column density at $4.0 \times 10^{21}$ cm$^{-2}$. This is roughly an order of magnitude lower than the initial column density of $5.0 \times 10^{22}$ cm$^{-2}$.

In order to  apply an objective procedure for characterizing the streams,  we employed the data clustering algorithm DBSCAN (part of the Python library {\it scikit-learn}, \citealt{scikit-learn}) to categorize the streams in our LTS model. DBSCAN views clusters as areas of high density (i.e., particle concentrations) separated by areas of low density. There are two parameters to the DBSCAN algorithm: \textit{min\_samples}, which represents the minimum number of points required to form a dense region, and \textit{eps} ($\epsilon$), which determines the minimum number of particles in a neighborhood for that neighborhood to be considered a ``point''. The algorithm starts with an arbitrary starting point. This point's $\epsilon$-neighborhood is explored, and if it contains sufficiently many points, a cluster is started. Otherwise, the point is labeled as noise. Note that this point might later be found in a sufficiently sized $\epsilon$-environment of a different point and hence be made part of a cluster. 

We built a routine using DBSCAN and applied it to the set of particle positions that have a density of at least $10^{5}$ cm$^{-3}$, in order to capture the densest parts of the streams. We set \textit{eps}, which determines the neighborhood radius of a ``point'' in a cluster, to $0.05$ pc, and \textit{min\_samples} to $10$.  All particles within the inner $2$ parsecs of the simulation were removed for this analysis, because the high-density region at the disk's inner edge formed due to the interaction of the gas with the artificial radial outflow. Therefore this high density region is ignored for our purposes.

We applied our DBSCAN routine to the LTS-t$5$-$50$ simulation at different times and were able to identify streams in all cases. These streams are transient, with an approximate timescale dictated by the turbulence injection interval ($0.1-0.2$ ~Myrs). The shape and quantity of the streams undergo frequent changes throughout the simulation; each stream is unique in terms of its structure and longevity, a consequence of the stochastic nature of the turbulence injections. Regardless, the constant energy injections revitalize the shearing perturbations that maintain general stream morphology within the model; these streams dissipate within $1-2$~Myrs if the injections cease.

\begin{figure*}
\centering
\includegraphics[width=0.45\textwidth]{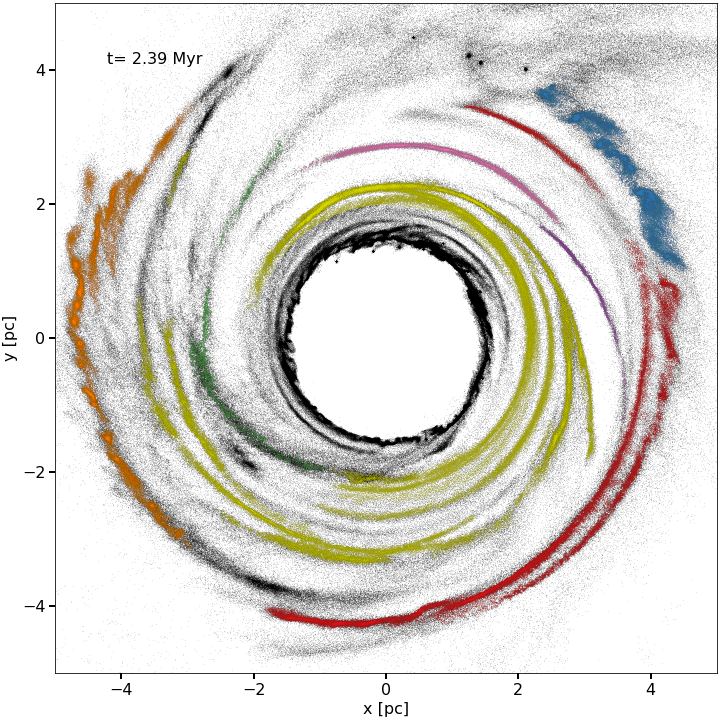}
\includegraphics[width=0.45\textwidth]{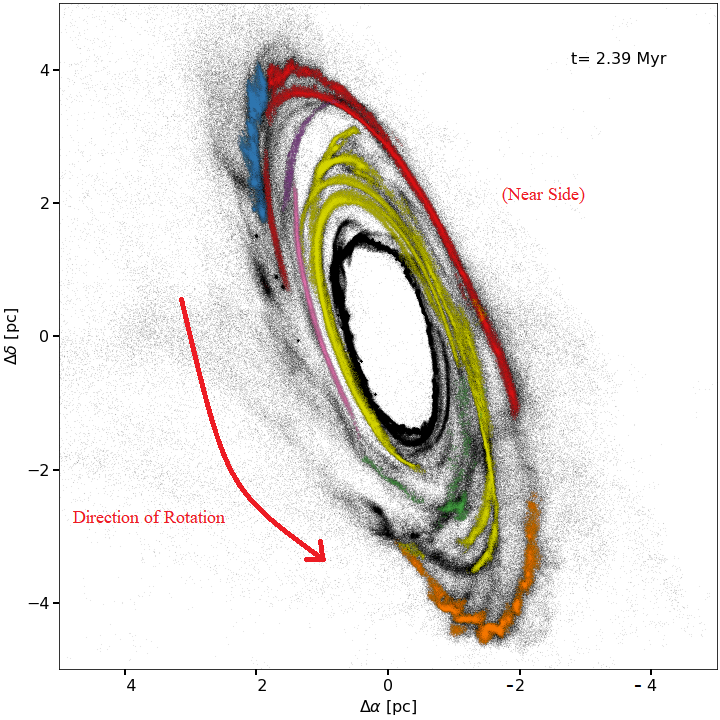}
\includegraphics[width=0.8\textwidth]{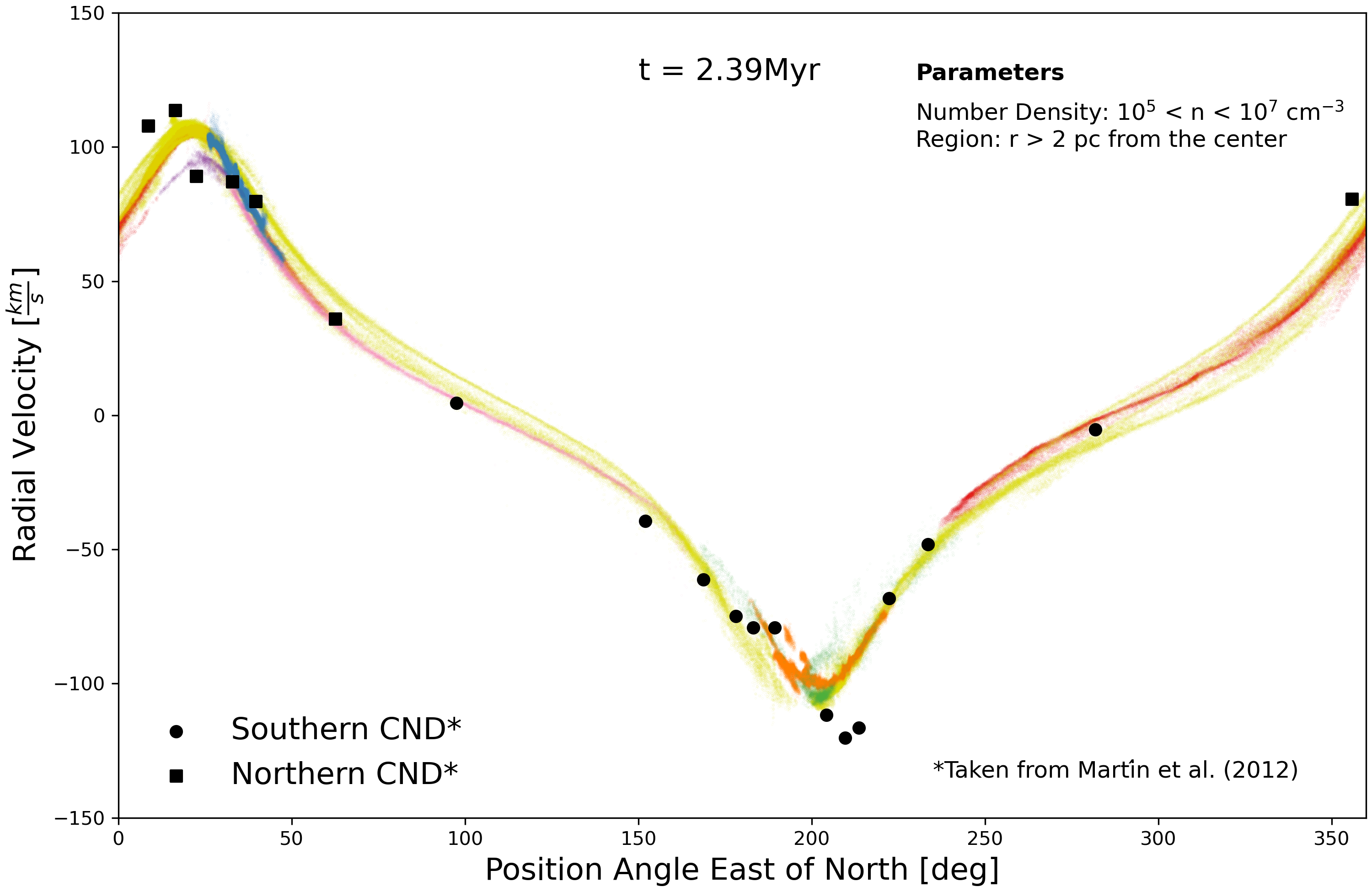}
\caption{Results from employing DBSCAN on the LTS-t$5$-$50$ model (\eturb = $10^{50}$ ergs with $\Delta t$ = 10$^{5}$ years) at 2.39 Myrs. The upper left panel displays seven identified streams (colored) imposed on the entire disk model (face-on view). The upper right panel accounts for the observed orientation. We do not consider the particles within a radius of $2$~pc in the stream finding analysis. We also note here that that the streams were first identified and colored in the face-on view before the model was rotated, thus they are less evident in the rotated orientation. The bottom panel shows the plot of radial velocities vs position angle of the particles in their corresponding streams (using the same colors). We limit the densities of the particles to be between $n=10^5$ and $n=10^7$ cm$^{-3}$. This trend closely follows the data points from \citet{Martin2012A&A...539A..29M}, which are reproduced in the Figure: square and circular points correspond to measurements made in the northern and southern CND, respectively.}
\label{fig:239_comp_streams}
\end{figure*}

A representative example of this analysis is shown in Figure \ref{fig:239_comp_streams}, where we plot the face-on (top left) and orientation-adjusted (top right) particle position plots of the LTS-t$5$-$50$ model at 2.39 Myrs. The number and characteristics of streams found at these times vary, but streams are always present and are easily identified by DBSCAN, as shown in different colors in Figure \ref{fig:239_comp_streams}. Furthermore, in the bottom panel of Figure \ref{fig:239_comp_streams} we plot the radial velocity vs position angle of the streams, as was done by \cite{Martin2012A&A...539A..29M} (c.f. their figure 9). The streams closely follow the trend outlined by the data points from their study, which were reproduced in the figure. However, there is much overlap between streams which would be difficult to discern if the streams were not colored. For example, at position angle $\sim200\degree$, there is noticeable overlap between the orange and yellow streams. It is possible that similar structures seen in CND observations may present a similar degeneracy, especially if observed structures have been categorized by visual inspections. Thus, we stress the importance of a clustering algorithm to find structures in the CND most effectively.

However, the DBSCAN routine was less effective at identifying streams when it was applied to the orientation-adjusted view because the tilt of the disk decreases particle separations and superposes streams upon each other, as shown in the upper right panel of Figure \ref{fig:239_comp_streams}. In addition, streams that happened to connect at one or more locations are often unidentifiable as separate entities. 

Consequently, our results indicate that there could be more streams than might normally be observed due to the disk’s orientation or the resolution. Using an enhanced version of DBSCAN or some equivalent algorithm, perhaps one that includes velocity information, such as {\it astrodendro} \citep{Rosolowsky2008ApJ...679.1338R}, could, in future work, facilitate the identification of streams and other structures in our models.

\section{Discussion}\label{sec:discussion}

In this paper, we have presented $3$D SPH simulations that explore the effects of turbulence on the dynamical evolution of the CND. These simulations, which include self-gravity and an adapted turbulence driving mechanism, are based on two models representing different driving scales. 

The Small Turbulence Scale (STS) models did not show sufficient structure to match the morphology revealed by observations (see Figure \ref{fig:STS_aden}). This is not surprising since the scales at which turbulence is stirred in this model are quite small ( $7.8\times10^{-4}$ - $0.05$ ~pc) and their effect on the much larger scale CND is only to endow it with a grainy appearance. This implies that sources of turbulence that operate on such small scales only have a minor impact on the overall dynamics of the CND. Sources that operate on these scales include stellar winds from Wolf-Rayet and massive main-sequence stars, AGB star winds, novae, pulsar wind nebulae, and magnetorotational instabilities (MRIs) from the predominantly torodial, few-milligauss magnetic field threading through the CND \citep[e.g.,][C.D. Dowell et al. in preparation]{Werner+88,Hildebrand+90,Hildebrand1993ApJ...417..565H}.

In contrast, we found that turbulence driving in the Large Turbulence Scale (LTS) models, which include the effects of larger scale ($0.03125$ - $2$ pc) sources of turbulence, such as supernova blast waves, have a much greater effect on the disk's evolution (see Figure \ref{fig:LTS_t550_aden_evol}). Here, the large-scale turbulence perturbations lead to the formation of streams and apparent clumps, some of which resemble observational features of the CND. However, these streams are partially obfuscated when we account for both the disk’s orientation and the general resolution of observations. As such, our results indicate that, not only can we match the streams observed in the CND, but there could be more streams than have been observed.

In addition, accounting for the disk's projection gives rise to density enhancements that appear as clumps (see Figure ~\ref{fig:convolve}). Further convolving the model to match the resolution of observations reinforces the apparent clump-like structure.  Such ``clumps'' are not evident in either the linear or log-scale column number density maps of the face-on model. This suggests that at least some observed clumps could be artifacts of both the resolution and the orientation of the CND. We posit that the CND is subject to episodic disturbances on large scales that, when sheared, leave a persistent pattern of spiral stream segments. In the absence of continued episodic injections of turbulence, the disk returns to a smooth state on a time scale of a few million years.

\subsection{Origin and Longevity of the CND}

The results of our LTS model show that streams and apparent clumps both naturally arise in a long-lived disk subjected to large-scale perturbations. Thus, tidally unstable clumps do not necessarily indicate that the disk itself is a transient feature, as has often been argued \citep[e.g.,][]{Guesten1987ApJ...318..124G, Requena-Torres2012A&A...542L..21R}. 
Our LTS simulations show that these clumps and streams are dynamic, constantly appearing and dissipating due to the injected turbulence.
Simulations by \cite{Blank2016MNRAS.459.1721B} also show that interactions between the nuclear stellar cluster's strong winds and the orbiting, annular disk cause instabilities at the inner edge of the CND. The existence of such mechanisms for continuously creating tidally unstable and therefore short-lived clumps and streams implies that such features cannot be used to conclude either that the CND is a transient feature or that the clumps within it must be dense enough ($n_{H_2}>3 \times 10^{7}$ cm$^{-3}$) to be virialized and therefore tidally stable, as has often been argued \citep[e.g.,][]{Vollmer:2001a, Vollmer:2001b, Shukla2004ApJ...616..231S,  Christopher2005ApJ...622..346C,Montero-Castano2009ApJ...695.1477M,Hsieh+21}.  Such high densities would imply extremely large CND masses of $10^5 - 10^6$ M$_{\odot}$, which are inconsistent with the optically thin far-IR and submillimeter fluxes from the CND \citep{Genzel2010RvMP...82.3121G, Etxaluze+11}. In addition, these studies assume that the molecular excitation is entirely collisional, whereas radiative excitation via the rotation-vibration lines is likely to contribute substantially \citep{Mills2013ApJ...779...47M}, thereby reducing the implied density.  Furthermore,
the  densities required for tidal stability are also much higher than have been inferred in many recent studies \citep{Oka+11, Requena-Torres2012A&A...542L..21R, Lau2013ApJ...775...37L, Mills2013ApJ...779...47M, SmithWardle14, Harada+15,Goicoechea+18a,  Tsuboi+18_ALMA_CND}.  We conclude that clumps and streams in the CND are most likely to be transient features, but that doesn't imply that the CND itself is a transient structure.  

Based on the observation that the CND appears to be disequilibrated because of the presence of internal transient structures, many publications model the formation of the CND as a recent event resulting from the infall of a dense cloud toward the central black hole \citep{Sanders1998MNRAS.294...35S, Bradford2005ApJ...623..866B, Bonnel2008Sci...321.1060B, Wardle2008ApJ...683L..37W, HobbsNayakshin09,  Alig2011MNRAS.412..469A, Oka+11,  Mapelli2012ApJ...749..168M, Mapelli+Trani16, Trani2018ApJ...864...17T, Goicoechea+18a,  Ballone+19}.  
The idea rests on the assumption that a relatively massive cloud with a very small total angular momentum can somehow be produced near the black hole, or can be produced further away, but be unimpeded in its accelerating trajectory toward the black hole.  However, those assumptions are questionable.  Although there remains some uncertainty about the line-of-sight placement of clouds in the central molecular zone (CMZ), gaseous structures in the CMZ appear to move on orbits that conform to the sense of Galactic rotation \citep[e.g.,][]{Henshaw+16}, even if they are influenced by a non-axisymmetric potential or by radial accelerations caused by extreme accretion or starburst activity at the center. Therefore CMZ clouds have considerable angular momentum. Creating a cloud with near-zero angular momentum, or one with a retrograde orbit that could eliminate the angular momentum of other clouds by colliding with them \citep[e.g.,][]{HobbsNayakshin09, TartenasZubovas20}, would therefore require it to be scattered at a large angle, but there are no known massive perturbers that could do that, including other clouds, because the sizes of clouds are large enough that they cannot approach each other closely enough to cause scattering by more than about $10^{\circ}$ without colliding and merging, thereby preserving their angular momentum.  Even with some undefined scattering process at work, the phase-space volume into which clouds would need to be scattered to be brought from a distance of more than several parsecs to the vicinity of the black hole is very small; the centrifugal barrier for a CMZ cloud scattered at almost any angle would be encountered at a radius larger than that of the CND\footnote{Supernova blast waves can push ambient gaseous material from moderately large distances into the zone of the CND \citep{Palous+20}, but the mass of material moved in that way is much smaller than the mass of the CND, and in any case, the initial angular momentum of the swept-up material imposes a centrifugal barrier on material swept inward by supernovae.}.  Furthermore, any large cloud having a non-conforming velocity would collide within a few dynamical times with other clouds and would be forced eventually into conformity with the CMZ.  Consequently, the production of the CND by a radially infalling cloud appears to be extremely difficult. We therefore conclude that the CND does not need to be interpreted as a transient feature, and that it can be an enduring structure that is fed piecemeal and relatively slowly and quasi-continuously by tidal streams from CMZ clouds already orbiting in its vicinity, as a number of authors have suggested \citep{Ho+91, Wright+01, Liu+12, Hsieh2017ApJ...847....3H, Tsuboi+18_ALMA_CND}.   Indeed, as noted by \citet{Tress+20}, the inherent shearing and stretching of gas in the CMZ, coupled with large-scale shocks that the gas encounters where $x1$ and $x2$ orbits intersect, and where CMZ gas interacts with gas migrating inward along the bar, causes the gas to assume a predominantly filamentary morphology.  In their simulations, \citet{Tress+20} find that the branched filamentation, caused in their model by dynamical feedback from star formation, extends all the way down to the CND, and that the CND is fed by mass inflow along such filamentary gas parcels.  These gaseous strands are not "clouds" {\em per se}, but they must carry the specific angular momentum of the gaseous structures in which they originated and of any swept-up gas that they have encountered.  Those with lesser amounts of angular momentum join the CND at radii at least as large as their centrifugal barrier. However, whether such gas parcels might plunge nearly undisturbed towards the center, as speculated by \citet{Tress+20}, remains to be demonstrated. Recent simulations by \citet{Salas2020arXiv201004170S,Salas2021AJ....161..243S}  show that turbulence injected into gas in the CMZ (for example by supernovae accompanying star formation) does lead to a steady inward migration of interstellar gas, but not to radially infalling clouds or streams. Furthermore, \citet{Palous+20} and Barna et al. (in preparation) show that individual supernovae within several parsecs of a galactic center can drive migration of gas into the central parsec around a supermassive black hole, but not to retrograde or radially infalling clouds or streams.

A hybrid model that invokes an infalling cloud colliding with a preexisting CND has been proposed by \citet{TartenasZubovas20}, who investigate how such an event might cause sufficient angular momentum loss of disk gas to provoke a major accretion event onto the central black hole. They find that the infalling cloud would need to have a retrograde orbit in order to remove enough angular momentum to cause a small-scale accretion disk to form and produce a high rate of accretion.  However, the challenge of producing retrograde orbits is similar to, if not worse than, that of producing near-zero angular momentum orbits. Retrograde orbits might be occasionally attainable for low-mass gas blobs, but seem unlikely for clouds with sufficient heft to disturb the CND in the required manner.  It is interesting to note, however, that the cloud impacts on the CND modelled by \citet{TartenasZubovas20} produce streams in the CND that are similar to those that are provoked by large-scale turbulence in our LTS model.

The evolution of the CND is also linked to its propensity to occasionally form stars and to its relationship to the central black hole.  The young nuclear cluster of massive stars occupying the central 0.5 pc presumably formed in a starburst event from an earlier manifestation of the CND, when the inner radius extended all the way in to the black hole and probably fed the black hole at a far higher rate than it is fed at present. This could happen again if the CND continues to be sustained by inwardly migrating material from the outside and if it undergoes viscous evolution that eventually causes its inner edge to move inwards. Indeed, this could be a repetitive process in which the CND undergoes a limit cycle of activity punctuated in each cycle by a relatively brief starburst event \citep{Morris1999AdSpR..23..959M}.  
To address such a scenario, future simulations are needed to account for the effects of the magnetic field, CND feeding from the outside, and other physical processes not included in our models. Additionally, continued investigations with ALMA and other high-resolution instruments will be invaluable for investigating the internal disk morphologies proposed for the CND in this paper.

\section*{Acknowledgments}

This material is based upon work supported by the National Science Foundation Graduate Research Fellowship Program under Grant No. DGE-1144087. This work used computational and storage services associated with the Hoffman2 Shared Cluster provided by UCLA Institute for Digital Research and Education's Research Technology Group. This work also used the Extreme Science and Engineering Discovery Environment (XSEDE) Comet at the San Diego Supercomputer Center at UC San Diego through allocation TG-AST180051. XSEDE is supported by National Science Foundation grant number ACI-1548562. SN acknowledges the partial support of NASA grant No. 80NSSC20K0500 and thanks Howard and Astrid Preston for their generous support. We also acknowledge the support by the UCLA Physics and Astronomy Research Program 2020. 

{\bf Software:}
Figure \ref{fig:STS_aden} to \ref{fig:convolve} were done using the SPH visualization software \splash\ \citep{Price2007}. We used a modified version of the Gadget2 code \citep{Springel2005} which includes our turbulence method. The version of the code can be found at \href{https://github.com/jesusms007/CNDturbulence}{https://github.com/jesusms007/CNDturbulence}. Finally, we used the unsupervised machine learning algorithm DBSCAN, which is part of the Python library {\it scikit-learn} \citep{scikit-learn}.

\appendix

\section{Artificial outflow}\label{appen:wind}
We showed in \cite{Salas2021AJ....161..243S} that our  turbulence driving module enhances inward angular momentum transport of gas. To counteract the rapid filling of the CND cavity due to turbulence, we need to mimic the effects of the outflow from the NSC on the inner edge of the CND. \cite{Blank2016MNRAS.459.1721B} modelled the NSC's outflow by assuming a number density of $n_0=100$ \invcmcube\ and a speed of $v_0=700$ km/s, propagating radially outward from a starting radius of $r=0.5$~pc, and found that its interaction with the inner edge of the CND creates instabilities and a shock at their interface. However, here we only approximate the effects of the outflow on the CND gas. 

Due to the outflow, a parcel of gas at the inner edge of the disk will gain a radially outward velocity, but the gas in the disk will impede its progress. We assume ram pressure balance, such that the parcel's speed is constant. From the parcel's frame of reference, it experiences a ram pressure $P_{outflow} = \rho_{0} v_{0}^2$ from one side (where $\rho_0$ and $v_0$ are the mass density and speed of the outflow, respectively), and a ram pressure from the opposite direction, $P_{CND} = \rho_{CND} v_{CND}^2$ (where $\rho_{CND}$ is the mass density of the neighboring gas, which is assumed to be identical to the parcel's, and $v_{CND}$ is the parcel's speed as seen from the CND's frame of reference). The parcel's speed will then be,
\begin{equation}
    v_{CND} = \sqrt{\frac{\rho_{0}}{\rho_{CND}}} v_{0} =  \sqrt{\frac{\mu_0 n_{0}}{\mu_{CND} n_{CND}}} v_{0} \ ,
\end{equation}
where $\mu_0$ and $\mu_{CND}$ are the mean mass per particle of the outflow's ionized gas, and the molecular CND gas, respectively. The ratio $\sqrt{\mu_0/\mu_{CND}}$ is subsumed in the factor $\zeta$, described below.

We then convert this treatment to work with Gadget2. However, here we consider the outflow's density at $r=1.5$~pc, and since the outflow's density decreases with $1/r^2$, we thus adopt a value of $n_0 = 10$ \invcmcube. Furthermore, because we are not accounting for the additional hydrodynamical interaction between the outflow and the CND, a speed of $v_0=700$ km/s, which is much larger than the escape velocity, would send SPH particles flying out to infinity, causing numerical problems in the code. We simplify this problem by assuming that the largest speed this artificial outflow will add to an SPH particle (which happens when the SPH particle number density, $n_i$, is equal to the outflow's, i.e., $n_i=n_0$. For particles with $n_i<n_0$, no radial speed is added. This limit is acceptable since we find in our simulations that they are very few particles with $n_i<n_0$ throughout the disk) equals a fraction of the particle's escape speed, $\zeta v_{esc,i}$. This parameter $\zeta$ was adjusted in order to maintain a stable radius of the CND's edge. In our testing, we found $\zeta=0.15$ satisfied this condition. 

Thus, to mimic the NSC's outflow, every timestep we add a radially outward speed to every SPH particle, $i$, within a radius $r=1.5$ pc equal to:
\begin{equation}
    v_i(r) = \zeta \sqrt{\frac{n_{0}}{n_{_i}}} v_{esc,i}(r) \ ,
\end{equation}
where $v_i(r)$, $v_{esc,i}(r)$ and $n_{i}$ are values corresponding to a given SPH particle.

\section{Finding streams with DBSCAN}\label{appen:dbscan}

We applied DBSCAN on the LTS-t$5$-$50$ model at multiple times after $1.5$ Myr (i.e., after streams fully encapsulate the disk). We present here two additional examples: $1.75$ Myrs and $2.99$ Myr.

\begin{figure*}
    \centering
   
  \includegraphics[width=0.45\textwidth]{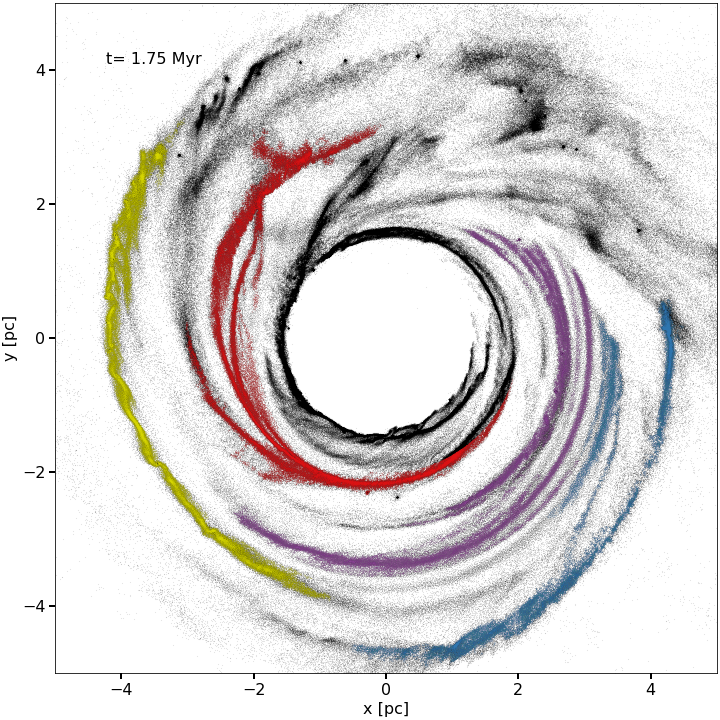}
  \includegraphics[width=0.45\textwidth]{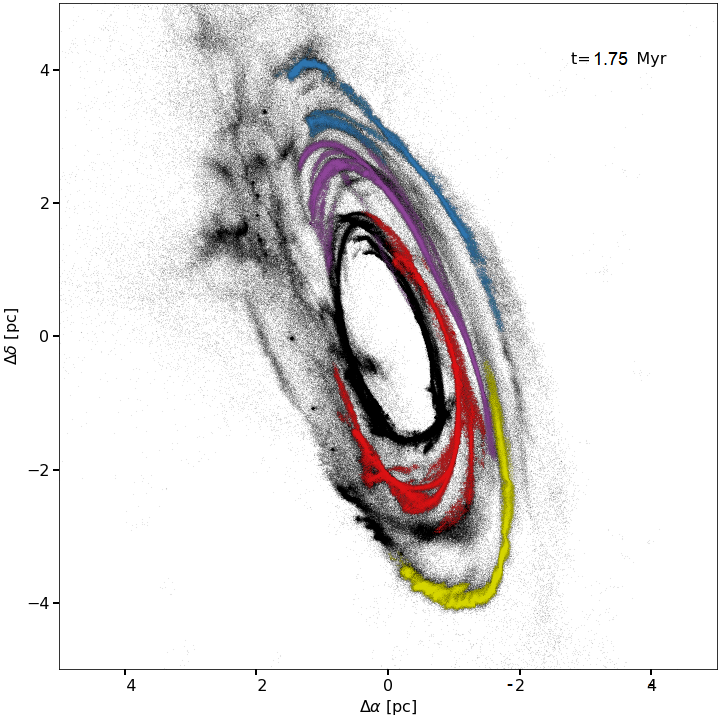}
    \includegraphics[width=0.8\textwidth]{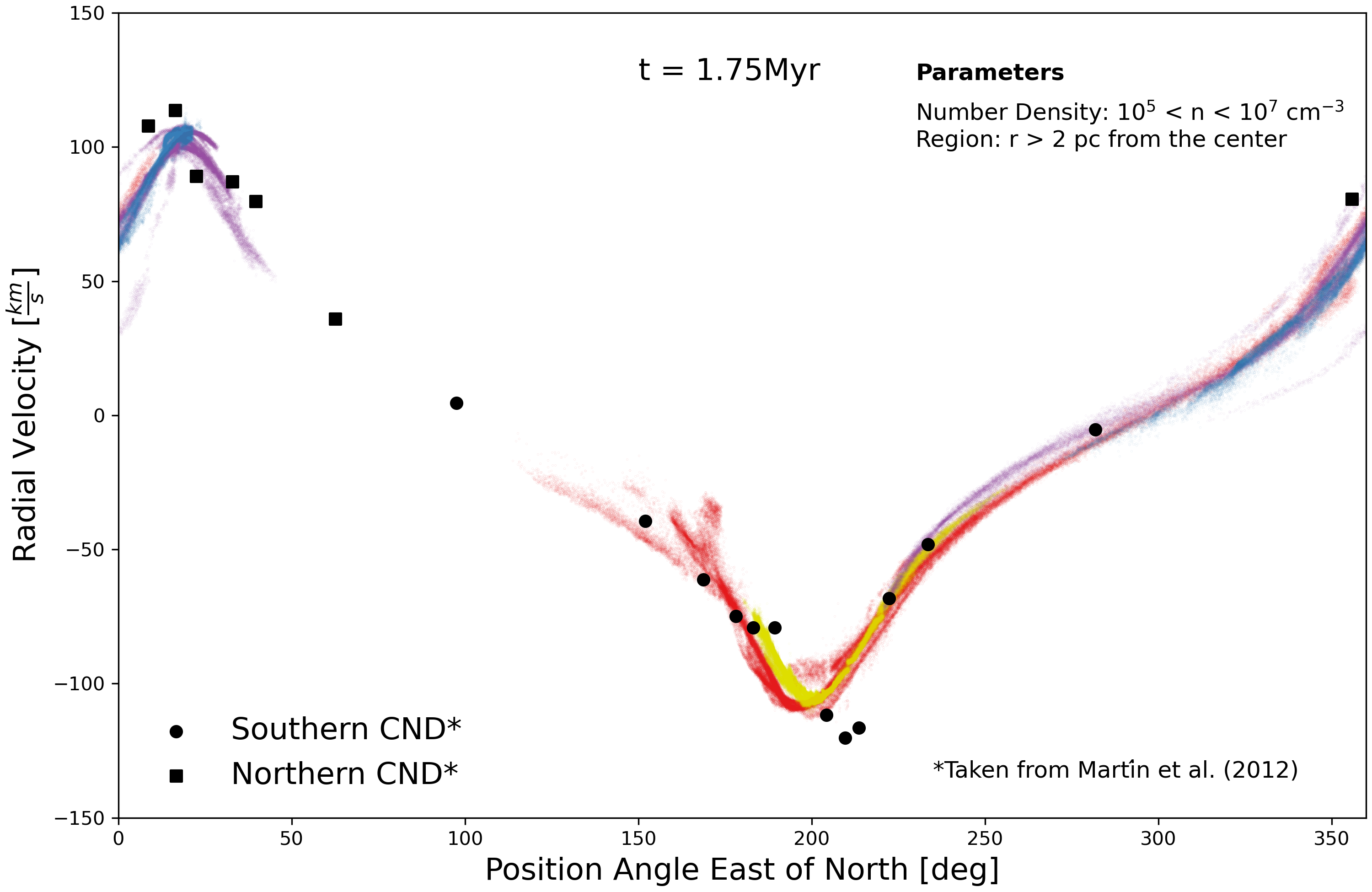}
    \caption{Results from employing DBSCANS on the LTS-t$5$-$50$ model (\eturb = $10^{50}$ ergs with $\Delta t$ = 10$^{5}$ years) before the run displayed in the results (1.75 Myrs). The upper left figure displays resulting four streams (colored) imposed on the entire disk model in the face on view. The upper right figure accounts for the rotation of the CND accounted for. The bottom figure maps the trend between the position angle and radial velocities of the particles in their corresponding streams (colored). This trend closely follows that found in \citealt{Martin2012A&A...539A..29M}, whose values have been fitted above. The square points correspond to what was realized as the Northern CND, whereas the circular points belong to the Southern CND.}
    \label{fig:175_comp_streams}
\end{figure*}

\begin{figure*}
    \centering
   
  \includegraphics[width=0.45\textwidth]{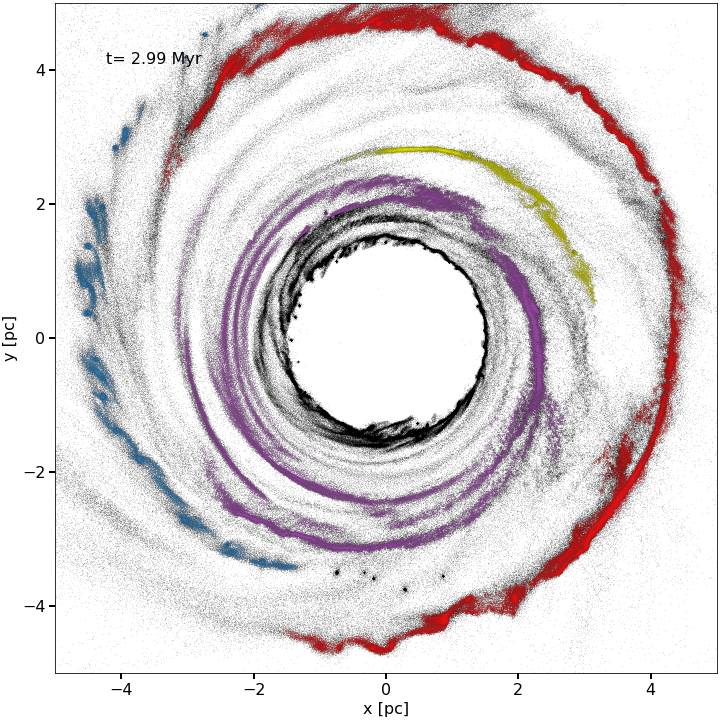}
  \includegraphics[width=0.45\textwidth]{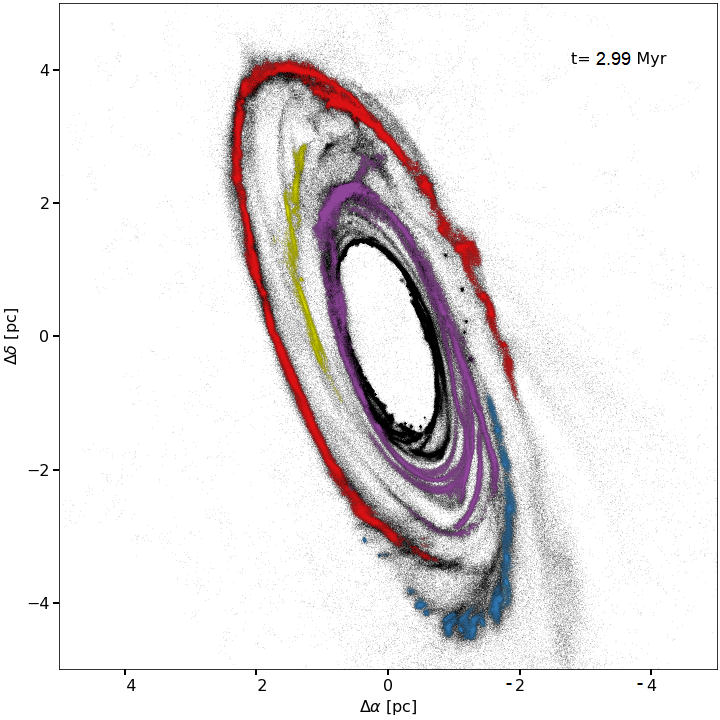}
    \includegraphics[width=0.8\textwidth]{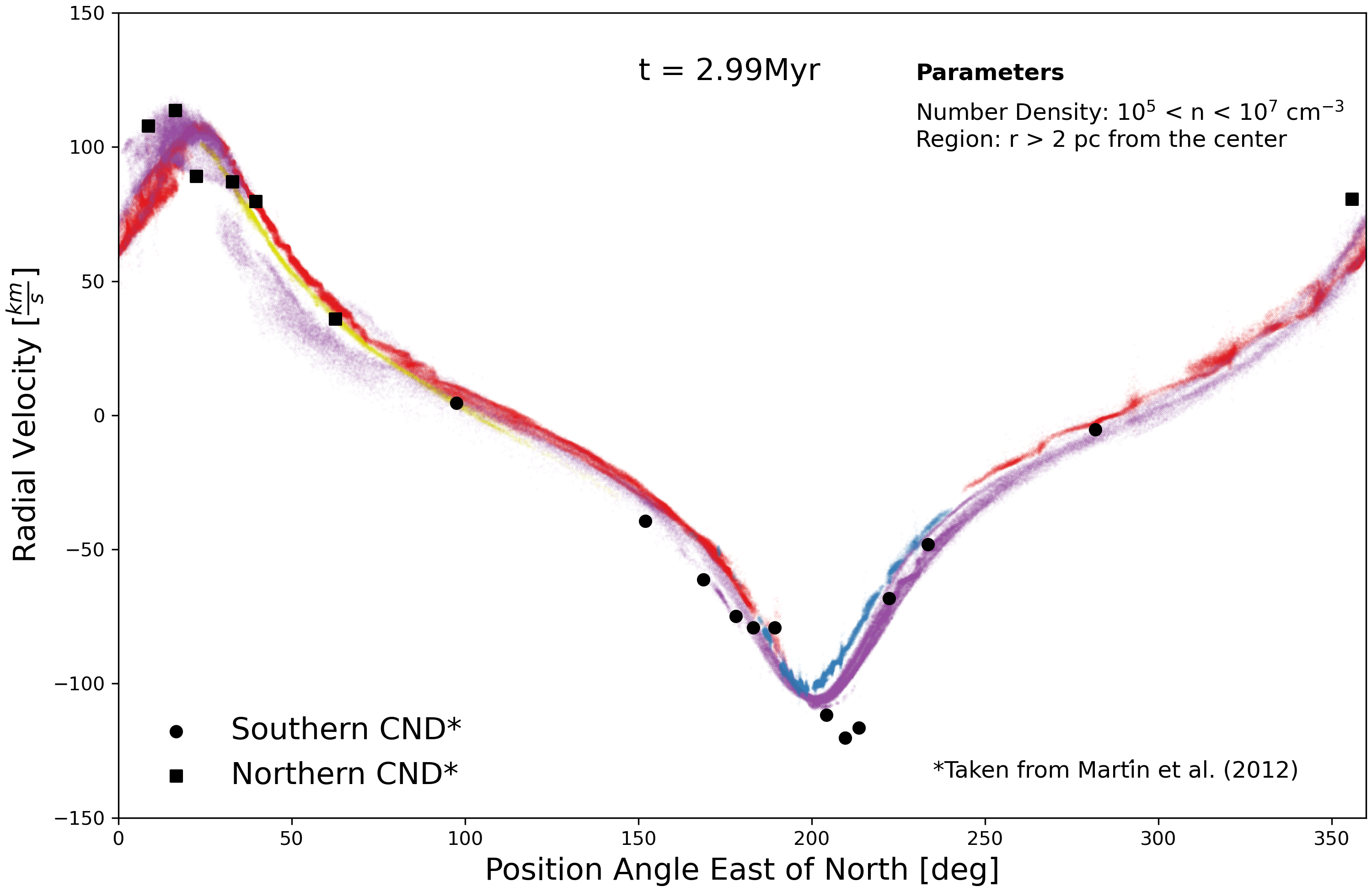}
    \caption{Results from employing DBSCANS on the LTS-t$5$-$50$ model (\eturb = $10^{50}$ ergs with $\Delta t$ = 10$^{5}$ years) after the run displayed in the results section (2.99 Myrs). The upper left figure displays resulting four streams (colored) imposed on the entire disk model in the face on view. The upper right figure is the same, but with the rotation of the CND accounted for. The bottom figure maps the trend between the position angle and radial velocities of the particles in their corresponding streams (colored). This trend closely follows that found in \citealt{Martin2012A&A...539A..29M}, whose values have been fitted above. The square points correspond to what was realized as the Northern CND, whereas the circular points belong to the Southern CND.}
    \label{fig:299_comp_streams}
\end{figure*}

\clearpage


\bibliographystyle{aasjournal}
\bibliography{references}

\end{document}